# Pan-Cancer Diagnostic Consensus Through Searching Archival Histopathology Images Using Artificial Intelligence

**Shivam Kalra**[1,2], **H.R. Tizhoosh**[*,2,3], **Sultaan Shah**[1], **Charles Choi**[1], **Savvas Damaskinos**[1], **Amir Safarpoor**[2], **Sobhan Shafiei**[2], **Morteza Babaie**[2], **Phedias Diamandis**[4], **Clinton JV Campbell**[5,6], and **Liron Pantanowitz**[7]

[1]Huron Digital Pathology, St. Jacobs, ON, Canada
[2]Kimia Lab, University of Waterloo, Waterloo, ON, Canada
[3]Vector Institute, MaRS Centre, Toronto, ON, Canada
[4]General Hospital/Research Institute (UHN), Toronto, Canada
[5]Stem Cell and Cancer Research Institute, McMaster University, Hamilton, Canada
[6]Department of Pathology and Molecular Medicine, McMaster University, Hamilton, Canada
[7]Department of Pathology, University of Pittsburgh Medical Center, PA, USA

## ABSTRACT

The emergence of digital pathology has opened new horizons for histopathology and related fields such as cytology. Computer programs and, in particular, artificial-intelligence algorithms, are able to operate on digitized slides to assist pathologists with diagnostic and theranostic tasks. Whereas machine learning involving classification and segmentation methods have obvious benefits for performing image analysis in pathology, image search represents an alternate and fundamental shift in computational pathology. Matching the pathology of new patients with already diagnosed and curated cases offers pathologist a novel and real-time approach to improve diagnostic accuracy through visual inspection of similar cases and computational majority vote for consensus building. In this study, we report the results from searching the largest public repository (The Cancer Genome Atlas [TCGA] program by National Cancer Institute, USA) of whole slide images from almost 11,000 patients depicting different types of malignancies. For the first time, we successfully indexed and searched almost 30,000 high-resolution digitized slides constituting 16 terabytes of data comprised of 20 million 1000x*1000 pixels image patches. The TCGA image database covers 25 anatomic sites and contains 32 cancer subtypes. High-performance storage and GPU power were employed for experimentation. The results were assessed with conservative "**majority voting**" to build consensus for subtype diagnosis through vertical search and demonstrated high accuracy values for both frozen sections slides (e.g., bladder urothelial carcinoma 93%, kidney renal clear cell carcinoma 97%, and ovarian serous cystadenocarcinoma 99%) and permanent histopathology slides (e.g., prostate adenocarcinoma 98%, skin cutaneous melanoma 99%, and thymoma 100%). The key finding of this validation study was that computational consensus appears to be possible for rendering diagnoses if a sufficiently large number of searchable cases are available for each cancer subtype.

## Introduction

Digital pathology is the virtual version of conventional microscopy utilized for the examination of glass pathology slides. In recent years, there has been accelerated adoption of digital pathology, whereby pathology laboratories around the world are slowly beginning to trade in their light microscopes for digital scanners, computers, and monitors. As a result, the pathology community has begun to scan many slides resulting in the creation of large databases of whole slide images (WSIs). The emergence of deep learning and other artificial-intelligence (AI) methods and their impressive pattern recognition capabilities when applied to these digital databases has immensely added to the value proposition of digital pathology[1–3]. Computerized operations, such as segmentation of tissue fragments and cell nuclei, and classification of diseases and their grades become possible after pathology slides are digitized. These operations could assist with many diagnostic and research tasks with expert-like accuracy when trained with the proper level of labeled data[4]. The majority of recent studies in digital pathology have reported the success of supervised AI algorithms for classification and segmentation[4–7]. This overrepresentation compared to other AI algorithms is related to the ease of design and in-lab validation to generate highly accurate results. However, compared to other methods of computer-vision algorithms, AI-based image search and retrieval offers a novel approach to computational

*Corresponding author: tizhoosh@uwaterloo.ca

pathology.

Content-based image search[8–11] implies that the input for search software is not text (e.g., disease description in a pathology report) but rather the input is an image such that the search and retrieval can be performed based on image pixels (visual content). Content-based image search is inherently unsupervised, which means that its design and implementation may not need manual delineation of a region of interest in the images[12–14]. More importantly, image search does not make any direct diagnostic decision on behalf of the pathologist; instead, it searches for similar images and retrieves them along with corresponding metadata (i.e., pathology reports), and displays them to the pathologist as decision support.

Variability in the visual inspection of medical images is a well-known problem[15–17]. Both inter- and intra-observer variability may affect image assessment and subsequently the ensuing diagnosis[18–21]. A large body of work have reported high rates of diagnostic inaccuracy as a result of major discordance among participating physicians with respect to case target diagnoses, and propose a combination of "routine second opinions" and "directed retrospective peer review"[22–24]. As most proposed AI-driven solutions for digital pathology mainly focus on the concept of classification, it appears that algorithmic decision-making may not necessarily contribute to supporting concordance by providing a framework for consensus building. Most capable classification schemes trained with immense effort are supposed to be used for triaging cases in the pathology laboratory and not for direct assistance in the pathologist'**s office[4]. In contrast, instantly retrieving multiple diagnosed cases with histopathologic similarity to the patient*'s biopsy about to be diagnosed offers a new generation of decision support that may even enable "virtual" peer review.

Content-based image retrieval (CBIR) systems have been under investigation for more than two decades[25–27]. Recently, deep learning has gained a lot of attention for image search[28–30]. Whilst CBIR systems of medical images have been well researched[11, 31–33], only with the emergence of digital pathology[34, 35] and deep learning[3, 36, 37] has research begun to focus on image search and analysis in histopathology [2, 38–40]. In the past 3 years, an image search engine called *Yottixel* has been designed and developed for application in pathology [32, 41–43]. Yottixel is a portmanteau for *one yotta pixel* alluding to the big-data nature of pathology images. The underlying technology behind Yottixel consists of a series of AI algorithms including clustering techniques, deep networks, and gradient barcoding. By generating a "bunch of barcodes" (BoB) for each WSI, digitized pathology slides can be indexed for real-time search. In other words, the tissue patterns of a WSI are converted into barcodes, a process that is both storage-friendly and computationally efficient. In this paper, we report the outcome of a comprehensive validation of the Yottixel search engine. We used WSI data from The Cancer Genome Atlas (TCGA) repository provided by the National Cancer Institute (NCI)/National Institutes of Health (NIH). Almost 30,000 WSI files of 25 primary anatomic sites and 32 cancer subtypes were processed by dismantling these large slides into almost 20,000,000 image patches (also called tiles) that were then individually indexed employing approximately 3,000,000 barcodes. This is the first time that the largest publicly available archive of WSIs has been employed to verify the performance of an image search engine for digital pathology.

## Methods

The Yottixel image search engine incorporates clustering, transfer learning, and barcodes and was used to conduct all experiments[30,32,41–47]. Before any search can be performed, all images in the repository have to be "indexed", i.e., every WSI is catalogued utilizing a "bunch of barcodes" (BoB indexing). These barcodes are stored for later use and generally not visible to the user. This process contains several steps (Figure 1):

- **Tissue Extraction** – Every WSI contains a bright (white) background that generally contains irrelevant (non-tissue) pixel information. In order to process the tissue, we need to *segment* the tissue region(s) and generate a black and white image (binary mask) that provides the location of all tissue pixels as "1" (white). Such a binary mask is depicted in the top row of Figure 1.

- **Mosaicking** – Segmented tissue now gets *patched* (divided into patches/tiles). These patches have a fixed size at a fixed magnification (e.g., 500×500 $\mu m^2$ at 20x scan resolution). All patches of the WSI get grouped into a pre-set number of categories (classes) via a clustering method (we used *k*-means algorithm[48]). A clustering algorithm is an unsupervised method that automatically groups WSI patches into clusters (i.e., groups) that contain similar tissue patterns. A small percentage (5%-20%) of all clustered patches are selected uniformly distributed within each class to assemble a **mosaic**. This mosaic represents the entire tissue region within the WSI. A sample mosaic consisting of 4 patches is depicted in the second row of Figure 1. Most WSIs we processed had a mosaic with around 70-100 patches.

- **Feature Mining** – All patches of the mosaic of each WSI are now pushed through pre-trained artificial neural networks (generally trained with *natural* images using datasets such as ImageNet[49]). The output of the network is ignored and the last pooling layers or the first connected layers are generally used as "features" to represent each mosaic patch. There could be approximately 1000-4000 features. The third row of Figure 1 shows this process where the features (colored squares) are passed on to the next stage, namely BoB indexing.

- **Bunch of Barcodes –** All feature vectors of each mosaic are subsequently converted into binary vectors using the *MinMax* algorithm[43]. This bunch of barcodes is the final index information for every query/input WSI that will be stored in the Yottixel index for future or immediate search. This is illustrated at the bottom of Figure 1.

In summary, Yottixel assigns "a bunch of barcodes" to each WSI to index the entire digital slide. The BoB indexing enables Yottixel to search a large archive of histopathology images very efficiently. The index can be easily shared among institutions if necessary. Technical details of Yottixel algorithms are described in a separate paper where its performance was tested with 2,300 WSIs[41].

Data Availability – We used publicly available image data as described in next section.

## Image Data

We used the publicly available dataset of 30,072 WSIs from the TCGA project[50, 51] (Genomic Data Commons GDC). We removed 952 WSIs due to the following reasons: poor staining, low resolution, lack of all magnification levels in the WSI pyramid, large presence of out-of-focus regions, and/or presence of unreadable regions within an image. In total, we processed 29,120 WSIs at 20x magnification (approximately 6 terabytes in compressed form) for this study. The dataset contains 25 anatomic sites with 32 cancer subtypes. Ten tumor types (brain, endocrine, gastrointestinal tract, gynecological, hematopoietic, liver/pancreaticobiliary, melanocytic, prostate/testis, pulmonary, urinary tract) had more than one primary diagnoses. From the 29,120 WSIs, 26,564 specimens were neoplasms, and 2,556 were non-neoplastic. A total of 17,425 files comprised of frozen section digital slides, and 11,579 files were of permanent hematoxylin and eosin (H&E) sections. For the remaining 116 WSIs, the tissue section preparation was unspecified. We did not remove manual pen markings from the slides when present. The TCGA codes for all 32 cancer subtypes are provided in Table 8 in the appendix. The TCGA dataset has a number of shortcomings[51]. Many of the cases are of frozen section in which tissue morphology may be compromised by frozen artifacts. Available cases may also reflect research bias in institutional biorepository collections. Furthermore, "tumours routinely subjected to neoadjuvant therapy may not have been able to be included in TCGA, because of limited availability of untreated specimens"[51]. Moreover, hematopathology is conspicuously absent from the TCGA dataset with just a few lymph nodes included. In spite of the shortcomings, the TCGA is the largest public dataset that can support a pan-cancer validation of AI solutions for digital pathology.

## Results

In two major series of experiments we calculated the "accuracy" of image search through "leave-one-patient-out" samplings. Whereas the literature of computer vision focuses on top-n accuracy (if any one of the n search results is correct, then the search is considered be to be successful), we calculated the majority-n accuracy (only if the majority among n search results were correct, the search was considered correct). Specifically, "correct" means that the tumor type (horizontal search) or tumor subtype within a specific diagnostic category (vertical search) was recognized correctly and matched by the majority of identified and retrieved cases. In order to avoid falsification of results through anatomic duplicates, we excluded all WSIs of the patient when one of the WSIs was the query.

### Horizontal Search: Cancer Type Recognition

The first series of experiments undertaken for all anatomic sites was *horizontal search*. The query WSI is compared against all other cases in the repository, regardless of anatomic site categorization. Of course, the primary anatomic site is generally known, and, in many cases, the cancer type may also be known to the pathologist. Thus, the purpose of the horizontal search (which is for either organ or cancer type recognition) is principally a fundamental algorithmic validation that may also have applications like searching for origin of malignancy in case of metastatic cancer.

The results of the horizontal search are depicted in Figure 2 (see appendix for details with Table 2 showing results for frozen section and Table 3 for permanent diagnostic slides). All experiments were conducted via "*leave-one-patient-out*" validation.

The following observations can be made from the results:

- Provided there are *sufficient* number of patients, we observed that the more we retrieve the more likely it was to achieve the right diagnosis: top-10 is better than top-5, and top-5 is better than top-3.

- General top-n accuracy that is common in the computer vision literature (top-3, top-5 and top-10 column in Tables 2 and 3) show high values but may not be suitable in the medical domain as it considers the search to be a success if at least one of the search results has the same cancer type as the query image.

- The majority vote among top n search results appears to be much more conservative and perhaps more appropriate as it only considers a search task as successful if the majority of top n search results show the same cancer type as the query image (majority-5 and majority-10 columns in Tables 2 and 3).

- With some exceptions, a general trend is observable that the more patients are available the higher the search-based consensus accuracy. The number of cases positively correlated with the majority vote accuracy for both frozen sections and permanent diagnostic slides.

**Vertical Search: Correctly Subtyping Cancer**

In the second series of experiments, we performed *vertical search*. Given the primary site of the query slide we confined the search only to WSIs from that organ. Hence, the goal of the vertical search was to recognize the cancer subtype. For this purpose, only those primary anatomic sites in the dataset with at least two possible subtypes were selected. Sample retrievals are illustrated in appendix Figure 8. The results for "leave-one-patient-out" validation are depicted in Figure 3 and Figure 4 (details in appendix, Table 4 for frozen sections and Table 5 for diagnostic slides).

Looking at the results of Figures 3 and 4 (Tables 4 and 5), we can observe the following:

- For both frozen sections and permanent diagnostic slides we continue to see a general trend whereby "*the more patients the better*" with both positive exceptions (KICH with 196 patients, and PCPG with 179 patients in Table 4) and negative exceptions (LUAD with 520 patients in Table 5).

- With majority-vote accuracy values for frozen sections (Table 4) in excess of 90% (KIRC, GBM, COAD, UCEC, PCPG), a search-based computational consensus appear to be possible when a large number of evidently diagnosed patients are available.

- With majority-vote accuracy values for diagnostic slides (Table 5) in excess of 90% (GBM, LGG, UCEC, KIRC, COAD, ACC, PCPG), a search-based computational consensus appear to be possible when a large number of evidently diagnosed patients are available.

- In most cases, it appeared that taking the majority of the top-7 search results provided the highest accuracy in most cases. However, the accuracy dropped drastically for subtypes with a small number of patients as we retrieved more and more images beyond 6 slides as the majority in such cases were taken from incorrect cases (we do not filter any result; no threshold is used; hence, all search results are considered as valid results).

- Based on all observations, it seems that there is a direct relationship between the number of diagnosed WSIs in the dataset and achievable consensus accuracy. For vertical search we calculated positive correlations of 0.5456 for frozen sections (Table 4) and 0.5974 for permanent diagnostic slides (Table 5). This trend was more pronounced for horizontal search with positive correlation of 0.7780 for frozen sections slides (Table 2) and 0.7201 for permanent diagnostic slides (Table 3).

- Additionally, the Cox-Stuart trend test[52] was used to check the upward monotonic trend of accuracy with respect to patients number. Having an increasing trend is considered as the null hypothesis for this test. The *p*-values for the horizontal (vertical) search are 1 (0.9991) and 0.9844 (0.9713) for frozen and diagnostic slides, respectively. Since the *p*-values are greater than the significance level (0.05), the null hypothesis is accepted. Consequently, there is a strong evidence of an upward monotonic trend.

Examining best, average, and worst cases for diagnostic slides, we randomly selected 3,000 slides and visualized them using the T-distributed Stochastic Neighbor Embedding (t-SNE) method[53] (see Figure 9). From this visualization we can observe that several subtype groups have been correctly extracted through search (see groups *a* to *f*). We can also observe the presence of outliers (e.g., DLBC in groups *a* and *b*). The outliers may be a product of the resolution of these scans, at least in part. At 20x magnification, for example, recognizing a diffuse large B-cell lymphoma (DLBC) from other large cell, undifferentiated non-hematopoietic tumors may not always be immediately possible for pathologists. This typically requires serial sections examined at multiple magnifications with ancillary studies such as immunohistochemistry.

**The Challenge of Validating Histologic Similarity**

One of the major benefits of using classification methods is that they can easily be validated; every image belongs to a class or not, a binary concept that can be conveniently quantified by counting the number of correctly/incorrectly categorized cases. In contrast, the concept of similarity in image search is intrinsically a fuzzy concept (i.e., cannot be answered with a simple yes/no in many cases) and mostly a matter of degree (*very similar*, *quite dissimilar*, etc.). Additionally, the similarity (or dissimilarity) between images is generally calculated using a distance metric/measure (in our case the Hamming distance[54]). The histologic similarity as perceived by pathologists may not correspond to tests where we used distance as a classification criterion. In other words, the classification-based tests that we run may be too harsh for search results and ignorant toward anatomic similarities among different organs.

One of the possible ways of examining the performance of the search is to look at the *heatmap*[55] of the confusion matrix. The values to construct the heatmap can be derived from the relative frequency of every subtype among the top 10 search results for a given subtype. A perfect heatmap would exhibit a pronounced diagonal with other cells being insignificant. Figure 5 shows the generated heatmap for all diagnostic subtypes in the dataset. The ordering of subtypes along the *y*-axis was done manually. It should be noted that our matching heatmap is not symmetrical like a correlation-based heatmap.

### *Analysis of the Heatmap*

The pronounced diagonal in Figure 5 shows that most disease subtypes have been correctly *classified* as they were very frequently retrieved among the top 10 horizontal search results. Other obvious observations:

- MESO is a difficult diagnosis with almost absent diagonal values.
- READ and COAD build a confusion region of 4 squares; they are confused with each other frequently.
- The same observation can be made for LUAD and LUSC. The vertical values for LUAD and LUSC also show that they are present in many other searches, for instance, when we search for UESC, HNSC and ESCA.
- LIHC is frequently among the search results for CHOL.
- For PRAD and BRCA we predominantly found PRAD and BRCA images, respectively.

Of note, the observational analysis of the heatmap alone may be limited. If we cluster (group) the search result frequencies and construct the dendrograms for the relationships in order to create an advanced heatmap, we might more easily discover the benefits of the search (see Figure 6). From there, we can observe:

- LGG and GBM are both glial tumors of the central nervous system
- Rectum and colon cancer are gland forming tumors of the colon
- Both uterine and ovarian carcinoma are grouped under gynecological
- Bladder, stomach and esophagus are upper gastrointestinal tumors
- Adenocarcinoma and squamous cell carcinoma are both subtypes of lung tumors
- Three kidney tumors appear close together

The errors (i.e., misclassifications) identified were still within the general grouping that the tumor originated from. Hence, from an image search perspective, it suggests that is it good at being close to the site of origin when it makes "classification" errors.

### **Chord Diagram of Image Search**

We used a chord diagram to further explore retrieved results. A chord diagram is the graphic display of the inter-relationships between numbers in a matrix. The numbers are arranged radially around a circle with the relationships between the data points generally visualized as arcs connecting the numbers/labels[56]. In Figure 7a, the chord diagram of horizontal search (cancer type recognition) for 11,579 permanent diagnostic slides of the TCGA dataset is illustrated. We can observe the following:

- Adenocarcinomas from several disparate organ systems match (e.g. colon, lung, stomach and breast). This is not surprising, as adenocarcinomas formed by glandular structures of equivalent grade in most organs are morphologically similar.
- Certain tumors derived from the same organ are related (e.g. LGG and GBM, UCEC and CESC, and Kidney RCC and KIRP).
- High-grade tumors from different anatomic locations appear to match (e.g. GBM and sarcoma). This may be attributed to the fact that such high-grade tumors likely display similar morphologic findings (e.g. necrosis).

- Squamous tumors from the head and neck and lung resemble urothelial carcinoma from the urinary bladder. In clinical practice this differential diagnosis can be morphologically challenging to diagnose, and thus warrants the use of ancillary studies such as immunohistochemistry to determine tumor origin.

- Hepatocellular carcinoma and thyroid carcinoma appear to exhibit the greatest number of matches (8 to 9) to other tumor subtypes. The significance of this finding is unclear.

- The broad relationship demonstrated among certain tumor subtypes is unexpected (e.g. cutaneous melanoma to sarcoma, LUSC and adenocarcinoma from several organs). Indeed, melanoma is known as the great mimicker in pathology given that these melanocytic tumors can take on many morphological appearances.

One has to emphasize that some relationships depicted in the chord diagram may disappear if distances are normalized and threshold applied. We did not filter any search results. No threshold was used. Hence, all search results were considered.

## Summary and Conclusions

The accelerated adoption of digital pathology is coinciding with and probably partly attributed to recent progress in AI applications in the field of pathology. This disruption in the field of pathology offers a historic chance to find novel solutions for major challenges in diagnostic histopathology and adjacent fields including biodiscovery. In this study, we indexed and searched the largest publicly available dataset of histopathology WSIs provided by the NIH/NCI. The question was whether one can build a computational consensus to potentially remedy the high intra- and inter-observer variability seen with diagnosing certain pathology tumors through search in a large archive of previously (and evidently) diagnosed cases. We performed a horizontal search to verify basic recognition capabilities of the image search engine. Furthermore, we performed leave-one-patient-out vertical searches to examine the accuracy of top *n* search results for establishing a diagnostic majority for cancer subtypes.

The results of this validation study show that building a computational consensus is possible if large and representative archives of well-characterized and evidently diagnosed cases are available. The ideal size of the dataset appears to be in excess of several thousand patients for each primary diagnosis and is most likely directly related to the anatomic complexity and intrinsic polymorphism of individual tissue types.

Future research should look into subtype consensus for individual primary diagnoses in more details for carefully curated datasets. As well, the need for much larger curated archives in the pathology community is clearly evident, which includes additional tissue types such as hematological. Lastly, comprehensive discordance measurement for subtypes with and without computational consensus should be planned and carried out as the ultimate evidence for the efficacy of the image search as a supportive diagnostic tool.

## References

1. Janowczyk, A. & Madabhushi, A. Deep learning for digital pathology image analysis: A comprehensive tutorial with selected use cases. *J. pathology informatics* **7** (2016).
2. Madabhushi, A. & Lee, G. Image analysis and machine learning in digital pathology: Challenges and opportunities (2016).
3. Tizhoosh, H. R. & Pantanowitz, L. Artificial intelligence and digital pathology: Challenges and opportunities. *J. pathology informatics* **9** (2018).
4. Campanella, G. *et al.* Clinical-grade computational pathology using weakly supervised deep learning on whole slide images. *Nat. medicine* **25**, 1301–1309 (2019).
5. Guo, Z. *et al.* A fast and refined cancer regions segmentation framework in whole-slide breast pathological images. *Sci. reports* **9**, 882 (2019).
6. Niazi, M. K. K., Parwani, A. V. & Gurcan, M. N. Digital pathology and artificial intelligence. *The Lancet Oncol.* **20**, e253–e261 (2019).
7. Xing, F., Xie, Y., Su, H., Liu, F. & Yang, L. Deep learning in microscopy image analysis: A survey. *IEEE transactions on neural networks learning systems* **29**, 4550–4568 (2017).
8. Lehmann, T. M. *et al.* Content-based image retrieval in medical applications. *Methods information medicine* **43**, 354–361 (2004).
9. Long, L. R., Antani, S., Deserno, T. M. & Thoma, G. R. Content-based image retrieval in medicine: retrospective assessment, state of the art, and future directions. *Int. J. Healthc. Inf. Syst. Informatics (IJHISI)* **4**, 1–16 (2009).
10. Markonis, D. *et al.* A survey on visual information search behavior and requirements of radiologists. *Methods information Medicine* **51**, 539–548 (2012).

11. Müller, H., Michoux, N., Bandon, D. & Geissbuhler, A. A review of content-based image retrieval systems in medical applications—clinical benefits and future directions. *Int. journal medical informatics* **73**, 1–23 (2004).

12. Sathya, R. & Abraham, A. Comparison of supervised and unsupervised learning algorithms for pattern classification. *Int. J. Adv. Res. Artif. Intell.* **2**, 34–38 (2013).

13. LeCun, Y., Kavukcuoglu, K. & Farabet, C. Convolutional networks and applications in vision. In *Proceedings of 2010 IEEE International Symposium on Circuits and Systems*, 253–256 (IEEE, 2010).

14. Onder, D., Sarioglu, S. & Karacali, B. Automated labelling of cancer textures in colorectal histopathology slides using quasi-supervised learning. *Micron* **47**, 33–42 (2013).

15. Elmore, J. G., Wells, C. K., Lee, C. H., Howard, D. H. & Feinstein, A. R. Variability in radiologists' interpretations of mammograms. *New Engl. J. Medicine* **331**, 1493–1499 (1994).

16. Mussurakis, S., Buckley, D., Coady, A., Turnbull, L. & Horsman, A. Observer variability in the interpretation of contrast enhanced mri of the breast. *The Br. journal radiology* **69**, 1009–1016 (1996).

17. Burnett, R. *et al.* Observer variability in histopathological reporting of malignant bronchial biopsy specimens. *J. clinical pathology* **47**, 711–713 (1994).

18. Winkfield, B., Aubé, C., Burtin, P. & Calès, P. Inter-observer and intra-observer variability in hepatology. *Eur. journal gastroenterology & hepatology* **15**, 959–966 (2003).

19. Louie, A. V. *et al.* Inter-observer and intra-observer reliability for lung cancer target volume delineation in the 4d-ct era. *Radiother. Oncol.* **95**, 166–171 (2010).

20. Cooper, W. A. *et al.* Intra-and interobserver reproducibility assessment of pd-l1 biomarker in non–small cell lung cancer. *Clin. Cancer Res.* **23**, 4569–4577 (2017).

21. Lewis Jr, J. S. *et al.* Inter-and intra-observer variability in the classification of extracapsular extension in p16 positive oropharyngeal squamous cell carcinoma nodal metastases. *Oral oncology* **51**, 985–990 (2015).

22. Peck, M., Moffat, D., Latham, B. & Badrick, T. Review of diagnostic error in anatomical pathology and the role and value of second opinions in error prevention. *J. Clin. Pathol.* **71**, 995–1000 (2018).

23. STROSBERG, C. *et al.* Second opinion reviews for cancer diagnoses in anatomic pathology: A comprehensive cancer center's experience. *Anticancer. research* **38**, 2989–2994 (2018).

24. Sasada, K. *et al.* Inter-observer variance and the need for standardization in the morphological classification of myelodysplastic syndrome. *Leuk. research* **69**, 54–59 (2018).

25. Veltkamp, R. C. & Tanase, M. Content-based image retrieval systems: A survey. *Dep. Comput. Sci. Utrecht Univ.* 1–62 (2002).

26. Singhai, N. & Shandilya, S. K. A survey on: content based image retrieval systems. *Int. J. Comput. Appl.* **4**, 22–26 (2010).

27. Zheng, L., Yang, Y. & Tian, Q. Sift meets cnn: A decade survey of instance retrieval. *IEEE transactions on pattern analysis machine intelligence* **40**, 1224–1244 (2017).

28. Babenko, A. & Lempitsky, V. Aggregating local deep features for image retrieval. In *Proceedings of the IEEE international conference on computer vision*, 1269–1277 (2015).

29. Liu, H., Wang, R., Shan, S. & Chen, X. Deep supervised hashing for fast image retrieval. In *Proceedings of the IEEE conference on computer vision and pattern recognition*, 2064–2072 (2016).

30. Kieffer, B., Babaie, M., Kalra, S. & Tizhoosh, H. R. Convolutional neural networks for histopathology image classification: Training vs. using pre-trained networks. In *2017 Seventh International Conference on Image Processing Theory, Tools and Applications (IPTA)*, 1–6 (IEEE, 2017).

31. Rahman, M. M., Bhattacharya, P. & Desai, B. C. A framework for medical image retrieval using machine learning and statistical similarity matching techniques with relevance feedback. *IEEE transactions on Inf. Technol. Biomed.* **11**, 58–69 (2007).

32. Tizhoosh, H. R. Barcode annotations for medical image retrieval: A preliminary investigation. In *2015 IEEE International Conference on Image Processing (ICIP)*, 818–822 (IEEE, 2015).

33. Qayyum, A., Anwar, S. M., Awais, M. & Majid, M. Medical image retrieval using deep convolutional neural network. *Neurocomputing* **266**, 8–20 (2017).

34. Farahani, N., Parwani, A. V. & Pantanowitz, L. Whole slide imaging in pathology: advantages, limitations, and emerging perspectives. *Pathol Lab Med Int* **7**, 23–33 (2015).

35. Liu, Y. & Pantanowitz, L. Digital pathology: Review of current opportunities and challenges for oral pathologists. *J. Oral Pathol. & Medicine* **48**, 263–269 (2019).

36. LeCun, Y., Bengio, Y. & Hinton, G. Deep learning. *nature* **521**, 436 (2015).

37. Goodfellow, I., Bengio, Y. & Courville, A. *Deep learning* (MIT press, 2016).

38. Komura, D. & Ishikawa, S. Machine learning methods for histopathological image analysis. *Comput. structural biotechnology journal* **16**, 34–42 (2018).

39. Shi, X. *et al.* Supervised graph hashing for histopathology image retrieval and classification. *Med. image analysis* **42**, 117–128 (2017).

40. Komura, D. *et al.* Luigi: Large-scale histopathological image retrieval system using deep texture representations. *BioRxiv* 345785 (2018).

41. Kalra, S., Choi, C., Shah, S., Pantanowitz, L. & Tizhoosh, H. Yottixel – an image search engine for large archives of histopathology whole slide images. *Submitt. to Med. Image Analysis* (2019).

42. Kumar, M. D., Babaie, M. & Tizhoosh, H. R. Deep barcodes for fast retrieval of histopathology scans. In *2018 International Joint Conference on Neural Networks (IJCNN)*, 1–8 (IEEE, 2018).

43. Tizhoosh, H. R., Zhu, S., Lo, H., Chaudhari, V. & Mehdi, T. Minmax radon barcodes for medical image retrieval. In *International Symposium on Visual Computing*, 617–627 (Springer, 2016).

44. Chenni, W., Herbi, H., Babaie, M. & Tizhoosh, H. R. Patch clustering for representation of histopathology images. *arXiv preprint arXiv:1903.07013* (2019).

45. Tizhoosh, H. R. & Babaie, M. Representing medical images with encoded local projections. *IEEE Transactions on Biomed. Eng.* **65**, 2267–2277 (2018).

46. Tizhoosh, H. R. & Czarnota, G. J. Fast barcode retrieval for consensus contouring. *arXiv preprint arXiv:1709.10197* (2017).

47. Tizhoosh, H. R., Mitcheltree, C., Zhu, S. & Dutta, S. Barcodes for medical image retrieval using autoencoded radon transform. In *2016 23rd International Conference on Pattern Recognition (ICPR)*, 3150–3155 (IEEE, 2016).

48. Jain, A. K. Data clustering: 50 years beyond k-means. *Pattern recognition letters* **31**, 651–666 (2010).

49. Deng, J. *et al.* Imagenet: A large-scale hierarchical image database. In *2009 IEEE conference on computer vision and pattern recognition*, 248–255 (Ieee, 2009).

50. Tomczak, K., Czerwińska, P. & Wiznerowicz, M. The cancer genome atlas (tcga): an immeasurable source of knowledge. *Contemp. oncology* **19**, A68 (2015).

51. Cooper, L. A. *et al.* Pancancer insights from the cancer genome atlas: the pathologist's perspective. *The J. pathology* **244**, 512–524 (2018).

52. Cox, D. R. & Stuart, A. Some quick sign tests for trend in location and dispersion. *Biometrika* **42**, 80–95 (1955).

53. Maaten, L. v. d. & Hinton, G. Visualizing data using t-sne. *J. machine learning research* **9**, 2579–2605 (2008).

54. Bookstein, A., Kulyukin, V. A. & Raita, T. Generalized hamming distance. *Inf. Retr.* **5**, 353–375 (2002).

55. Wilkinson, L. & Friendly, M. The history of the cluster heat map. *The Am. Stat.* **63**, 179–184 (2009).

56. Holten, D. Hierarchical edge bundles: Visualization of adjacency relations in hierarchical data. *IEEE Transactions on visualization computer graphics* **12**, 741–748 (2006).

## Acknowledgements

We would like to thank the Ontario Government for awarding an ORF-RE grant for this project (Ontario Research Fund - Research Excellence). The first author is supported by an MITACS internship. The basic research of the corresponding author leading to this work has been supported by The Natural Sciences and Engineering Research Council of Canada (NSERC) through multiple Discovery Grants. For access to computational and storage facilities, we like to thank Dell EMC and Teknicor for their generous support.

# Appendix

| TCGA Code | Primary Diagnosis | #Patients |
|---|---|---|
| ACC | Adrenocortical Carcinoma | 86 |
| BLCA | Bladder Urothelial Carcinoma | 410 |
| BRCA | Breast Invasive Carcinoma | 1097 |
| CESC | Cervical Squamous Cell Carcinoma and Endocervical Adenocarcinoma | 304 |
| CHOL | Cholangiocarcinoma | 51 |
| COAD | Colon Adenocarcinoma | 459 |
| DLBC | Lymphoid Neoplasm Diffuse Large B-cell Lymphoma | 48 |
| ESCA | Esophageal Carcinoma | 185 |
| GBM | Glioblastoma Multiforme | 604 |
| HNSC | Head and Neck Squamous Cell Carcinoma | 473 |
| KICH | Kidney Chromophobe | 112 |
| KIRC | Kidney Renal Clear Cell Carcinoma | 537 |
| KIRP | Kidney Renal Papillary Cell Carcinoma | 290 |
| LGG | Brain Lower Grade Glioma | 513 |
| LIHC | Liver Hepatocellular Carcinoma | 376 |
| LUAD | Lung Adenocarcinoma | 522 |
| LUSC | Lung Squamous Cell Carcinoma | 504 |
| MESO | Mesothelioma | 86 |
| OV | Ovarian Serous Cystadenocarcinoma | 590 |
| PAAD | Pancreatic Adenocarcinoma | 185 |
| PCPG | Pheochromocytoma and Paraganglioma | 179 |
| PRAD | Prostate Adenocarcinoma | 499 |
| READ | Rectum Adenocarcinoma | 170 |
| SARC | Sarcoma | 261 |
| SKCM | Skin Cutaneous Melanoma | 469 |
| STAD | Stomach Adenocarcinoma | 442 |
| TGCT | Testicular Germ Cell Tumors | 150 |
| THCA | Thyroid Carcinoma | 507 |
| THYM | Thymoma | 124 |
| UCEC | Uterine Corpus Endometrial Carcinoma | 558 |
| UCS | Uterine Carcinosarcoma | 57 |
| UVM | Uveal Melanoma | 80 |

**Table 1.** The TCGA codes (in alphabetical order) of all 33 primary diagnoses and corresponding number of evidently diagnosed patients in the dataset (TCGA = The Cancer Genome Atlas)

| Tumor Type | WSI Count | Patient Count | Top-10 | Top-5 | Top-3 | Majority-5 | Majority-10 |
|---|---|---|---|---|---|---|---|
| Brain | 1797 | 1083 | 97.44 | 95.21 | 92.76 | 82.24 | 83.86 |
| Gynecological | 2216 | 1450 | 97.60 | 93.50 | 88.22 | 67.96 | 68.86 |
| Pulmonary | 1634 | 1068 | 95.34 | 90.75 | 83.90 | 58.01 | 59.30 |
| Gastrointestinal tract | 1947 | 1212 | 95.12 | 87.98 | 81.86 | 61.32 | 62.86 |
| Breast | 1495 | 1075 | 93.44 | 88.56 | 83.87 | 65.61 | 66.35 |
| Prostate/testis | 755 | 634 | 91.92 | 87.28 | 84.63 | 66.22 | 68.07 |
| Urinary tract | 1980 | 1300 | 90.25 | 83.48 | 79.89 | 62.67 | 64.59 |
| Endocrine | 769 | 729 | 84.78 | 71.39 | 61.89 | 30.68 | 35.37 |
| Melanocytic malignancies | 532 | 529 | 83.83 | 68.79 | 57.51 | 25.93 | 29.13 |
| Liver, pancreaticobiliary | 659 | 602 | 81.48 | 73.29 | 63.73 | 30.34 | 35.35 |
| Hematopoietic | 181 | 169 | 78.45 | 73.48 | 69.06 | 44.19 | 45.85 |
| Head and neck | 663 | 465 | 70.88 | 57.16 | 48.11 | 22.32 | 26.24 |
| Mesenchymal | 259 | 255 | 56.37 | 42.85 | 33.59 | 06.17 | 11.19 |

**Table 2.** Results for cancer type recognition (**horizontal search**) among frozen slides. Every whole slide image was compared with all other slides in the repository regardless of the primary site. The table is sorted based on Top-10 accuracy numbers. (WSI= whole slide image)

| Tumor Type | WSI Count | Patient Count | Top-10 | Top-5 | Top-3 | Majority-5 | Majority-10 |
|---|---|---|---|---|---|---|---|
| Brain | 1692 | 870 | 98.99 | 97.81 | 96.69 | 91.37 | 91.60 |
| Pulmonary | 1109 | 1011 | 98.46 | 96.12 | 91.70 | 75.83 | 76.19 |
| Prostate/testis | 701 | 550 | 97.43 | 94.86 | 92.15 | 80.31 | 82.88 |
| Breast | 1116 | 1049 | 95.96 | 91.57 | 87.09 | 70.87 | 71.50 |
| Gastrointestinal tract | 1144 | 1108 | 95.54 | 90.73 | 85.83 | 65.12 | 67.91 |
| Urinary tract | 1374 | 1275 | 95.41 | 90.82 | 85.51 | 66.01 | 69.21 |
| Gynecological | 1039 | 933 | 95.28 | 90.37 | 84.50 | 63.71 | 66.89 |
| Endocrine | 936 | 732 | 94.55 | 91.88 | 88.67 | 73.93 | 77.13 |
| Liver, pancreaticobiliary | 618 | 585 | 93.85 | 87.37 | 82.20 | 63.75 | 64.72 |
| Head and neck | 466 | 446 | 90.55 | 82.40 | 75.96 | 49.14 | 54.50 |
| Melanocytic malignancies | 551 | 509 | 88.20 | 79.31 | 70.41 | 37.20 | 43.73 |
| Mesenchymal | 594 | 253 | 87.37 | 80.63 | 73.73 | 50.84 | 53.70 |
| Hematopoietic | 221 | 163 | 84.61 | 81.44 | 76.47 | 52.03 | 56.56 |

**Table 3.** Results for cancer type recognition (**horizontal search**) among diagnostic slides. Every whole slide image was compared with all other slides in the repository regardless of the primary site. The table is sorted based on Top-10 accuracy numbers. (WSI= whole slide image)

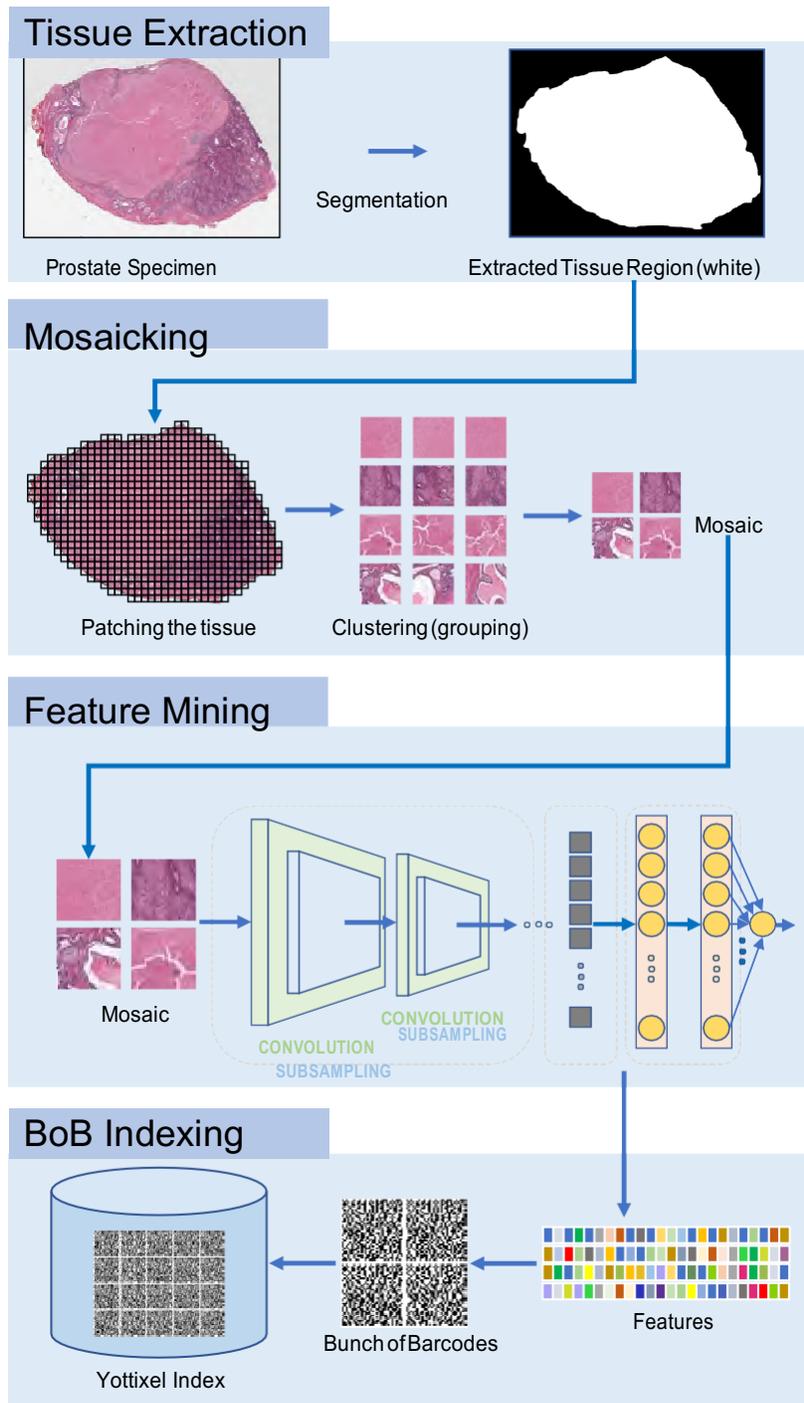

**Figure 1.** Yottixel Image Search Engine: Whole-slide images are segmented first to extract the tissue region by excluding the background (top block). A mosaic of representative patches (tiles) is assembled through grouping of all patches of the tissue region using an unsupervised clustering algorithm (second block from the top). All patches of the mosaic are fed into a pre-trained artificial neural network for feature mining (third block from the top). Finally, a bunch of barcodes is generated and added to the index of all WSI files in the archive (bottom block).

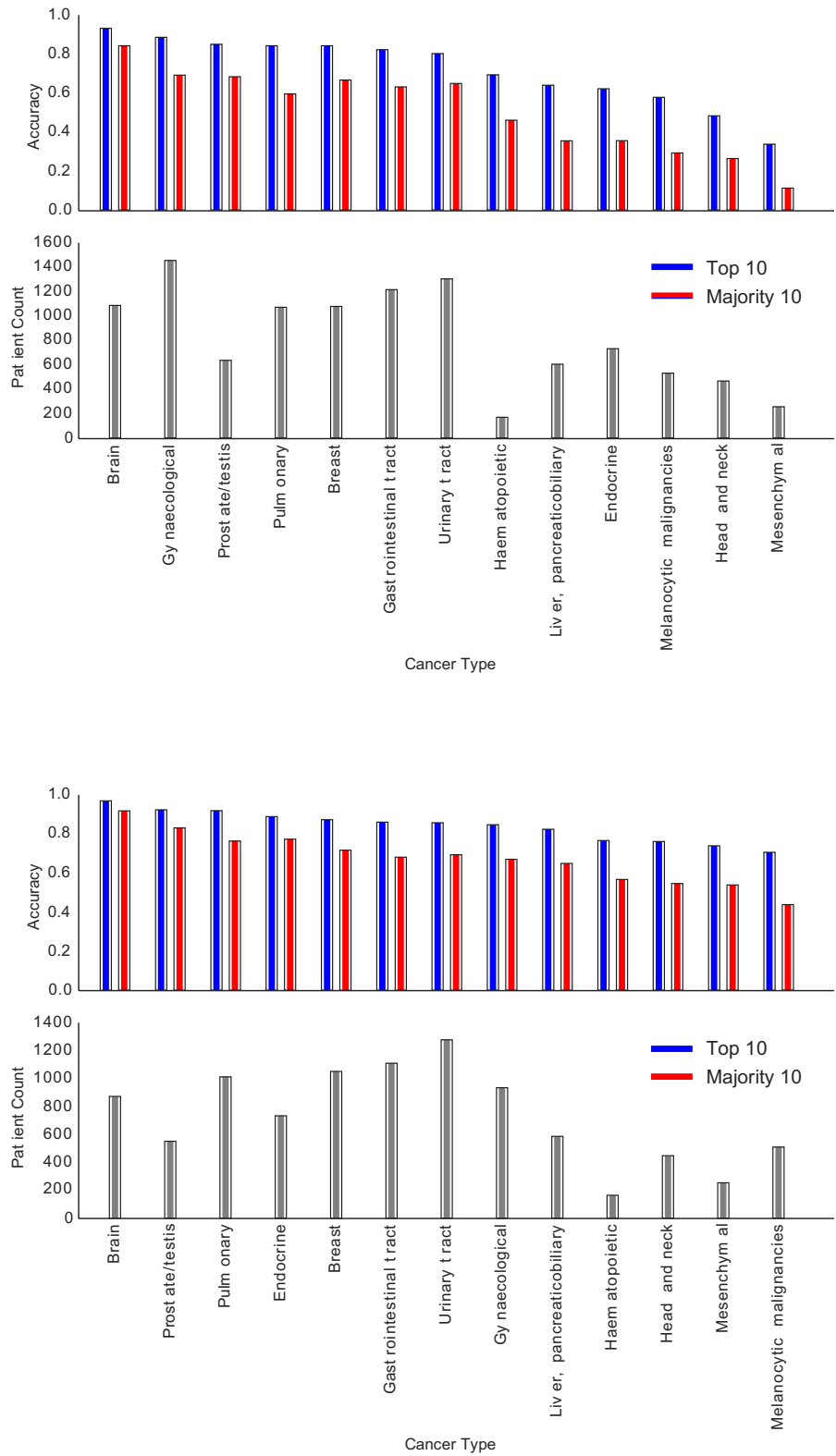

**Figure 2.** Horizontal search for frozen sections (top) and permanent diagnostic slides (bottom). Details are demonstrated in Tables 2 and 3 in the appendix.

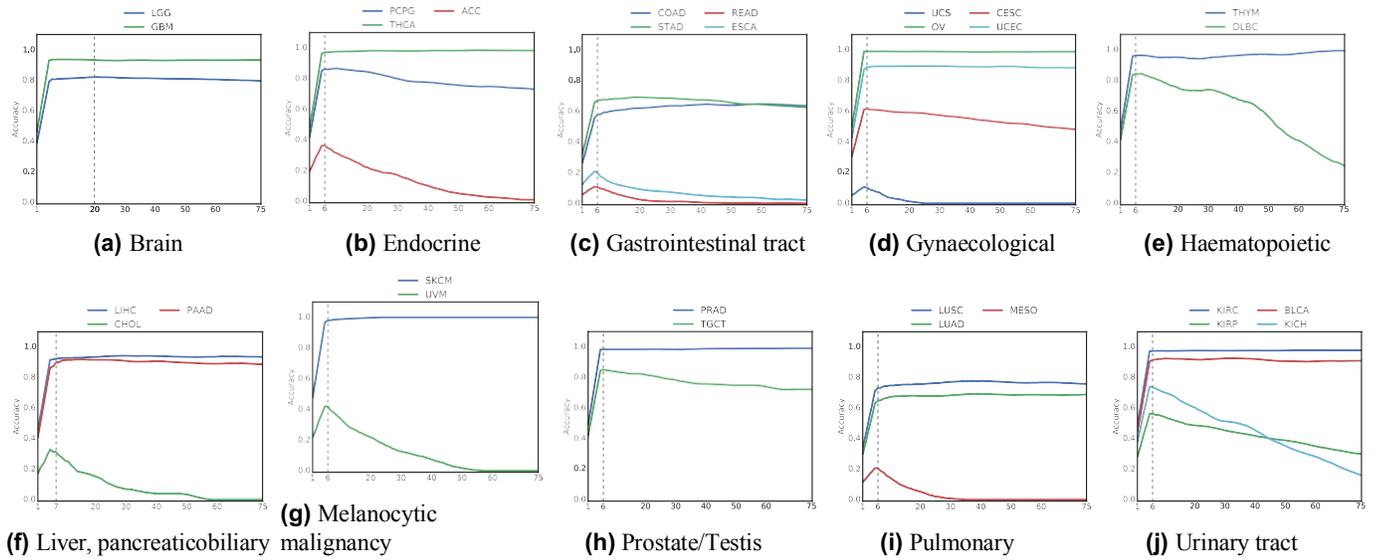

**Figure 3.** Vertical search in frozen sections slides from anatomic sites with at least two cancer subtypes.

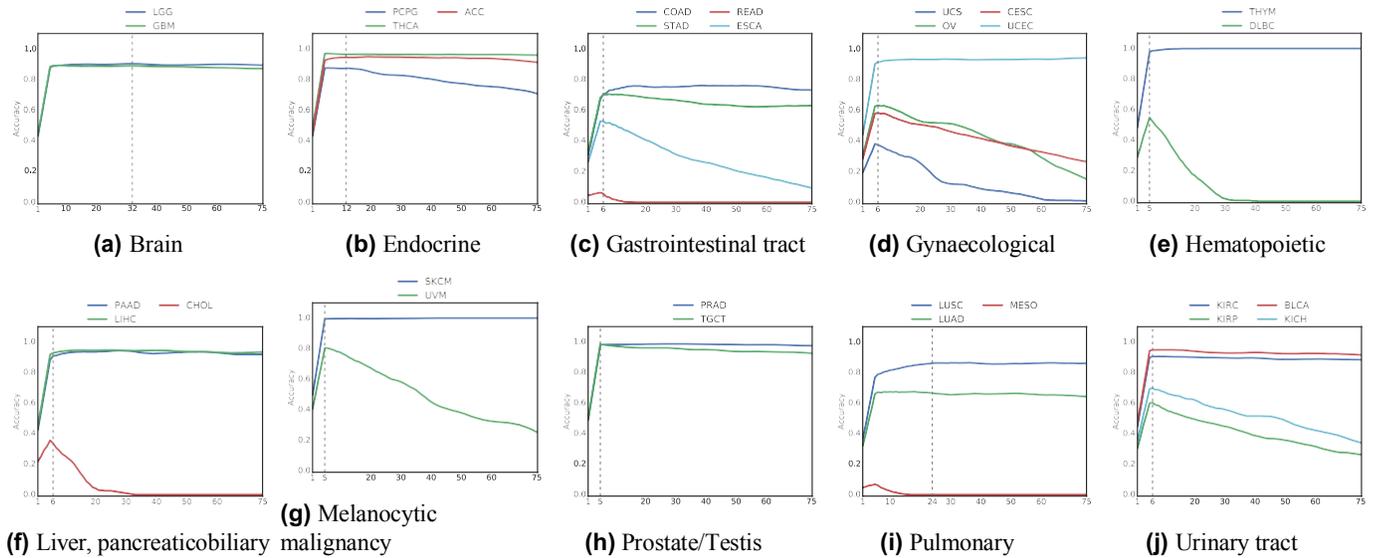

**Figure 4.** Vertical search in permanent diagnostic slides from anatomic sites with at least two cancer subtypes.

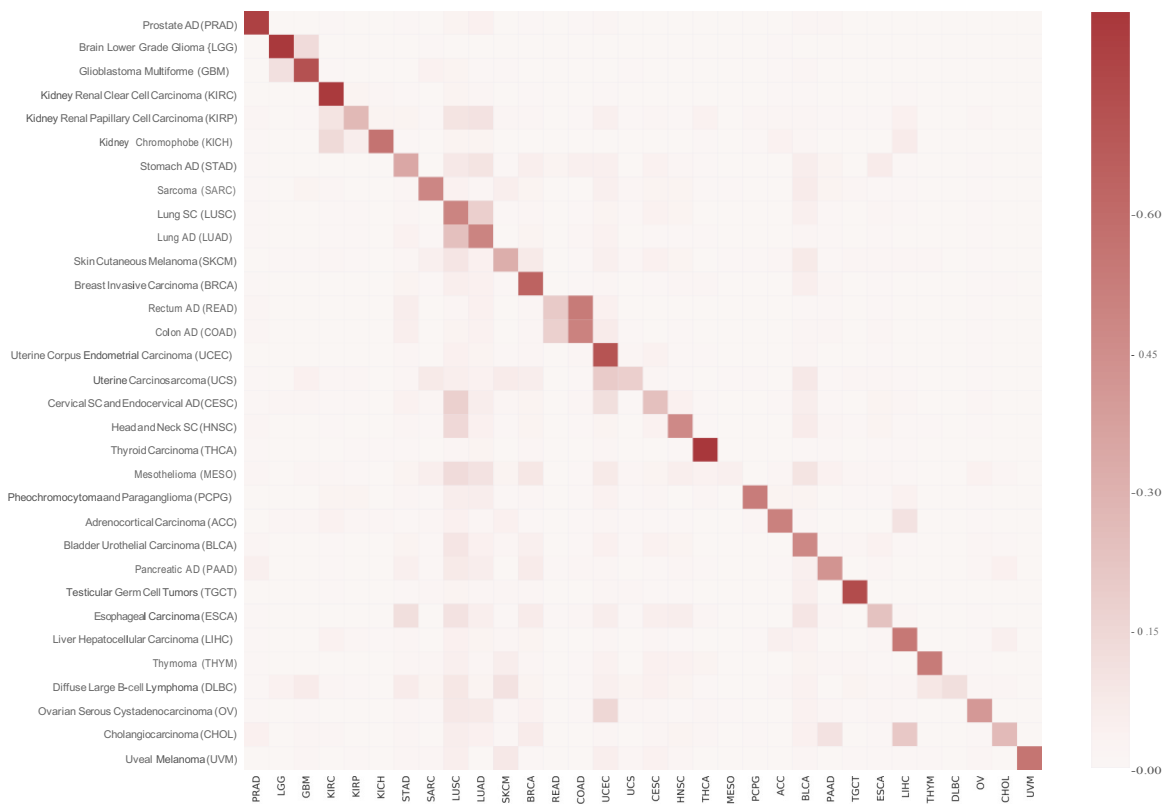

**Figure 5.** Heatmap of re-scaled relative frequency of matched (red) and mismatched (pale) search results for each diagnosis from permanent diagnostic slides. Re-scaling of frequencies was done through dividing each frequency by the total number of slides for each subtype.

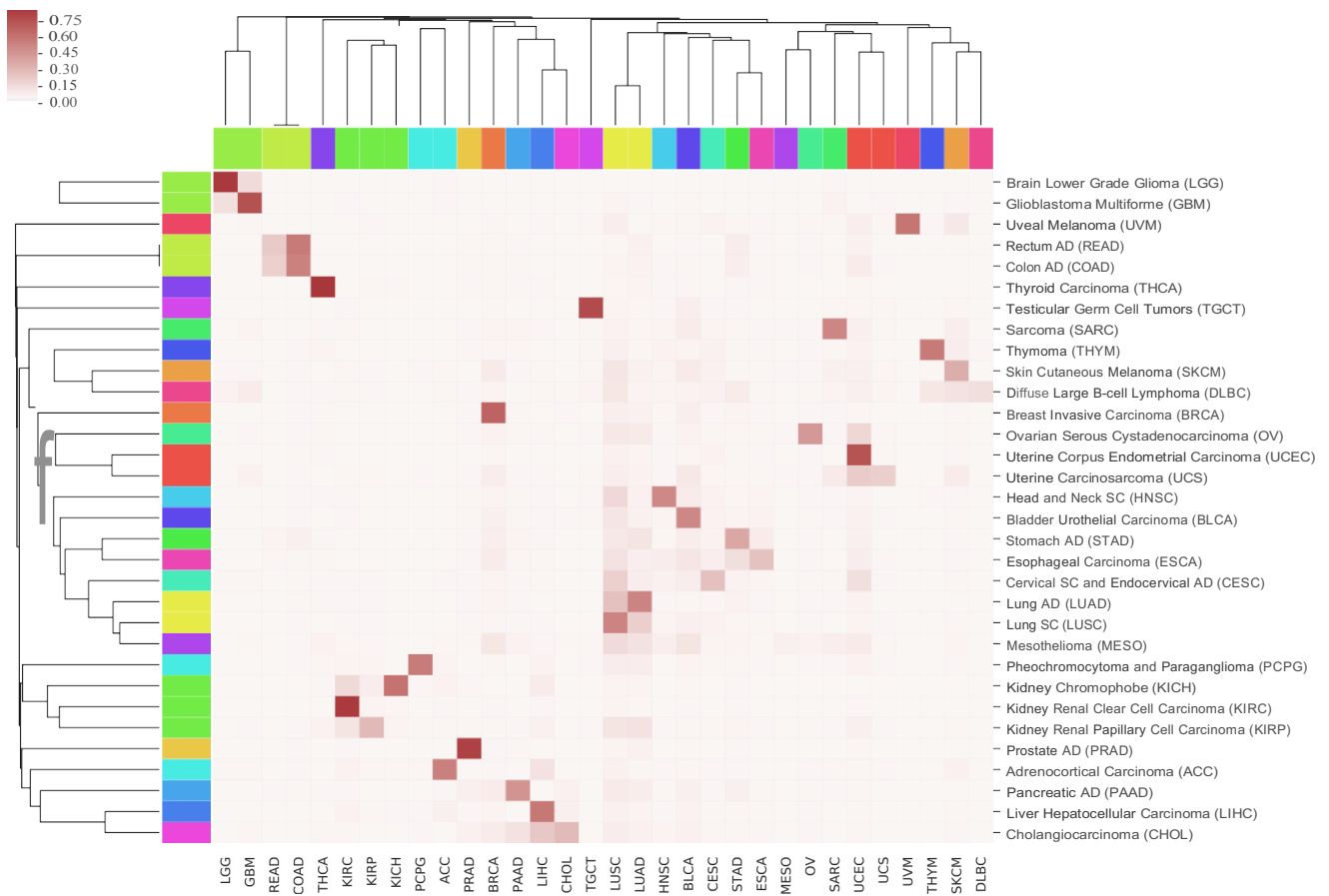

**Figure 6.** Dendrograms of clustered relative search frequencies.

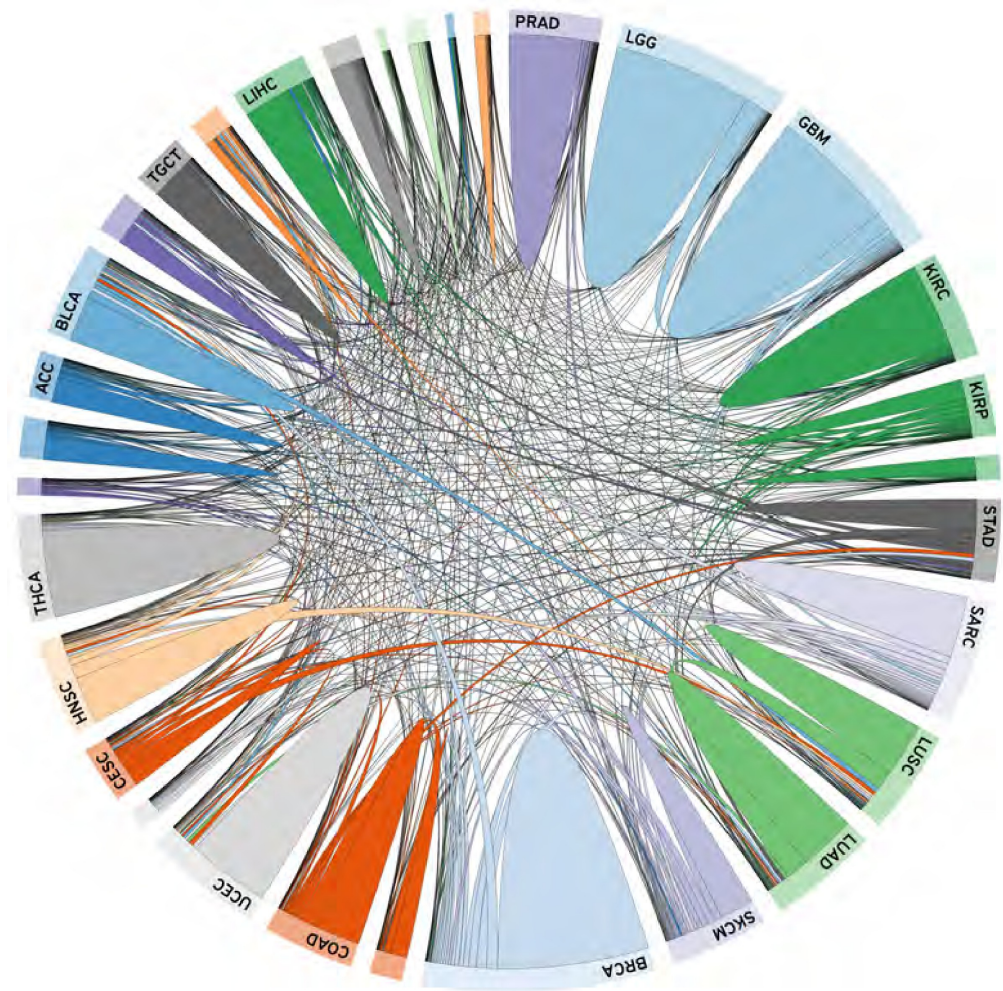

**(a)** Chord Diagram

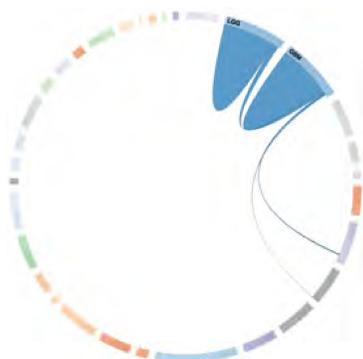

**(b)** Brain

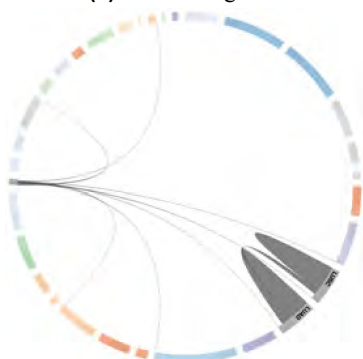

**(c)** Pulmonary

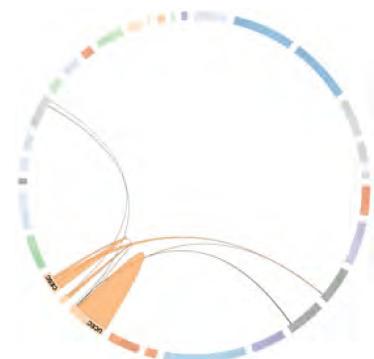

**(d)** Gynaecological

**Figure 7.** Chord diagram of horizontal image search for diagnostic slides of the TCGA dataset (a). Sample relations for brain (LGG and GBB), pulmonary (LAUD, LUSC and MESO) and gynecological (UCEC, UCS and CESC). The chord diagram can be interactively viewed online: https://bit.ly/2k6g3k1.

| Tumor Type | WSI Count | Patient Count | Majority-3 | Majority-5 | Majority-7 | Majority-10 | Majority-15 | Majority-20 |
|---|---|---|---|---|---|---|---|---|
| Brain | | | | | | | | |
| GBM | 1102 | 582 | 94.37 | 94.19 | 94.64 | 92.74 | 94.28 | 92.92 |
| LGG | 695 | 501 | 83.59 | 82.58 | 82.58 | 80.28 | 82.58 | 81.00 |
| Endocrine | | | | | | | | |
| ACC | 81 | 81 | 45.67 | 45.67 | 39.50 | 28.39 | 30.86 | 20.98 |
| PCPG | 174 | 170 | 88.50 | 85.63 | 87.35 | 86.20 | 85.63 | 83.90 |
| THCA | 514 | 478 | 97.08 | 97.08 | 97.27 | 97.47 | 97.66 | 97.85 |
| Gastrointestinal tract | | | | | | | | |
| COAD | 830 | 449 | 61.92 | 63.73 | 63.25 | 56.62 | 64.21 | 60.00 |
| ESCA | 166 | 165 | 33.73 | 25.90 | 19.27 | 12.04 | 12.65 | 09.03 |
| STAD | 623 | 428 | 69.98 | 71.10 | 70.46 | 65.48 | 70.94 | 67.41 |
| READ | 328 | 170 | 16.15 | 14.32 | 13.71 | 05.48 | 05.79 | 02.13 |
| Gynecological | | | | | | | | |
| OV | 1184 | 586 | 99.32 | 99.07 | 98.90 | 98.98 | 99.07 | 98.81 |
| CESC | 298 | 291 | 67.44 | 64.42 | 63.42 | 59.06 | 61.40 | 58.05 |
| UCS | 49 | 49 | 16.32 | 10.20 | 14.28 | 04.08 | 04.08 | 02.04 |
| UCEC | 685 | 524 | 89.78 | 90.07 | 90.94 | 89.05 | 90.21 | 89.34 |
| Hematopoietic | | | | | | | | |
| DLBC | 57 | 45 | 85.96 | 91.22 | 85.96 | 80.70 | 78.94 | 73.68 |
| THYM | 124 | 124 | 96.77 | 97.58 | 96.77 | 95.16 | 94.35 | 95.16 |
| Liver, pancreaticobiliary | | | | | | | | |
| LIHC | 392 | 370 | 92.85 | 93.36 | 93.36 | 92.60 | 93.36 | 93.62 |
| CHOL | 51 | 51 | 39.21 | 35.29 | 37.25 | 19.60 | 17.64 | 13.72 |
| PAAD | 216 | 181 | 89.35 | 91.66 | 92.59 | 90.74 | 92.12 | 90.74 |
| Melanocytic malignancies | | | | | | | | |
| SKCM | 463 | 460 | 98.05 | 98.70 | 98.48 | 98.48 | 99.35 | 99.56 |
| UVM | 69 | 69 | 55.07 | 46.37 | 47.82 | 31.88 | 28.98 | 18.84 |
| Prostate/testis | | | | | | | | |
| TGCT | 155 | 149 | 88.38 | 86.45 | 87.09 | 83.87 | 83.22 | 81.29 |
| PRAD | 600 | 485 | 98.83 | 98.33 | 98.50 | 98.33 | 98.33 | 98.50 |
| Pulmonary | | | | | | | | |
| LUSC | 745 | 485 | 76.37 | 78.25 | 77.98 | 70.87 | 77.18 | 73.42 |
| LUAD | 806 | 500 | 67.86 | 68.23 | 69.72 | 64.14 | 71.09 | 66.12 |
| MESO | 83 | 83 | 28.91 | 27.71 | 20.48 | 14.45 | 08.43 | 03.61 |
| Urinary tract | | | | | | | | |
| BLCA | 420 | 401 | 92.38 | 92.85 | 93.80 | 90.95 | 93.33 | 90.95 |
| KICH | 138 | 88 | 81.15 | 78.26 | 74.63 | 68.11 | 68.84 | 57.24 |
| KIRC | 1055 | 529 | 97.63 | 97.81 | 97.81 | 97.25 | 97.53 | 97.63 |
| KIRP | 367 | 282 | 63.48 | 62.12 | 60.76 | 51.22 | 52.58 | 47.13 |

**Table 4.** Results for cancer subtype identification (**vertical search**) among frozen section slides. Only those primary sites were considered for vertical search which had at least two subtypes in the repository. A positive correlation of 0.57 was measured between the number of patients and the highest accuracy.

| Tumor Type | WSI Count | Patient Count | Majority-3 | Majority-5 | Majority-7 | Majority-10 | Majority-15 | Majority-20 |
|---|---|---|---|---|---|---|---|---|
| Brain | | | | | | | | |
| GBM | 851 | 381 | 90.36 | 91.18 | 91.06 | 87.89 | 89.65 | 88.13 |
| LGG | 841 | 489 | 90.48 | 89.77 | 90.60 | 88.58 | 91.08 | 89.17 |
| Endocrine | | | | | | | | |
| ACC | 227 | 56 | 92.07 | 93.83 | 94.71 | 94.27 | 94.71 | 94.71 |
| PCPG | 196 | 176 | 89.79 | 88.77 | 88.77 | 85.71 | 88.77 | 84.18 |
| THCA | 513 | 500 | 97.85 | 97.66 | 97.27 | 96.68 | 96.49 | 96.49 |
| Gastrointestinal tract | | | | | | | | |
| COAD | 436 | 428 | 69.03 | 76.14 | 77.06 | 69.72 | 78.21 | 74.31 |
| ESCA | 157 | 155 | 62.42 | 59.87 | 59.23 | 45.22 | 47.77 | 39.49 |
| READ | 157 | 156 | 16.56 | 10.19 | 05.09 | 00.63 | 00.00 | 00.00 |
| STAD | 394 | 369 | 73.60 | 75.12 | 72.58 | 67.76 | 71.31 | 67.00 |
| Gynecological | | | | | | | | |
| UCEC | 566 | 505 | 91.69 | 92.22 | 93.10 | 91.69 | 93.99 | 92.75 |
| CESC | 277 | 267 | 62.09 | 62.45 | 62.45 | 54.51 | 54.15 | 49.09 |
| UCS | 90 | 56 | 42.22 | 42.22 | 40.00 | 32.22 | 32.22 | 27.77 |
| OV | 106 | 105 | 66.98 | 66.98 | 66.03 | 59.43 | 59.43 | 51.88 |
| Hematopoietic | | | | | | | | |
| DLBC | 43 | 43 | 67.44 | 58.13 | 58.13 | 37.20 | 34.88 | 16.27 |
| THYM | 178 | 120 | 98.87 | 98.87 | 98.87 | 99.43 | 100.00 | 100.00 |
| Liver, pancreaticobiliary | | | | | | | | |
| CHOL | 39 | 39 | 51.28 | 43.58 | 33.33 | 25.64 | 20.51 | 02.56 |
| LIHC | 378 | 364 | 92.32 | 93.65 | 94.17 | 93.65 | 94.97 | 94.44 |
| PAAD | 201 | 182 | 92.03 | 91.04 | 92.03 | 92.03 | 94.52 | 93.03 |
| Melanocytic malignancies | | | | | | | | |
| UVM | 80 | 80 | 85.00 | 83.75 | 82.50 | 77.50 | 73.75 | 68.75 |
| SKCM | 471 | 429 | 99.57 | 99.57 | 99.57 | 99.57 | 99.78 | 99.57 |
| Prostate/testis | | | | | | | | |
| TGCT | 254 | 149 | 100.00 | 99.21 | 98.03 | 96.85 | 96.45 | 96.06 |
| PRAD | 447 | 401 | 98.43 | 98.43 | 98.21 | 98.21 | 98.43 | 98.43 |
| Pulmonary | | | | | | | | |
| LUAD | 520 | 465 | 73.26 | 70.96 | 71.53 | 63.26 | 70.76 | 64.42 |
| MESO | 86 | 74 | 13.95 | 08.13 | 05.81 | 02.32 | 00.00 | 00.00 |
| LUSC | 503 | 472 | 78.92 | 81.70 | 83.69 | 78.13 | 85.28 | 83.30 |
| Urinary tract | | | | | | | | |
| BLCA | 454 | 384 | 95.37 | 95.81 | 95.59 | 94.27 | 95.37 | 93.61 |
| KIRC | 516 | 511 | 91.66 | 91.66 | 91.27 | 90.11 | 90.31 | 89.53 |
| KICH | 108 | 108 | 74.07 | 75.92 | 71.29 | 66.66 | 62.96 | 59.25 |
| KIRP | 296 | 272 | 68.91 | 67.22 | 63.17 | 53.04 | 54.05 | 48.31 |

**Table 5.** Results for cancer subtype identification (**vertical search**) among permanent diagnostic slides. Only those primary sites were considered for vertical search which had at least two subtypes in the repository. A positive correlation of 0.49 was measured between the number of patients and the highest accuracy.

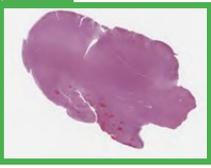

**Figure 8.** Sample retrievals for cancer subtype categorization through majority votes. The top four slides are of permanent diagnostic slides whereas the bottom three slides are of frozen section slides. The misclassified and successful queries are marked with red and green boundaries, respectively. (for abbreviations see Table )

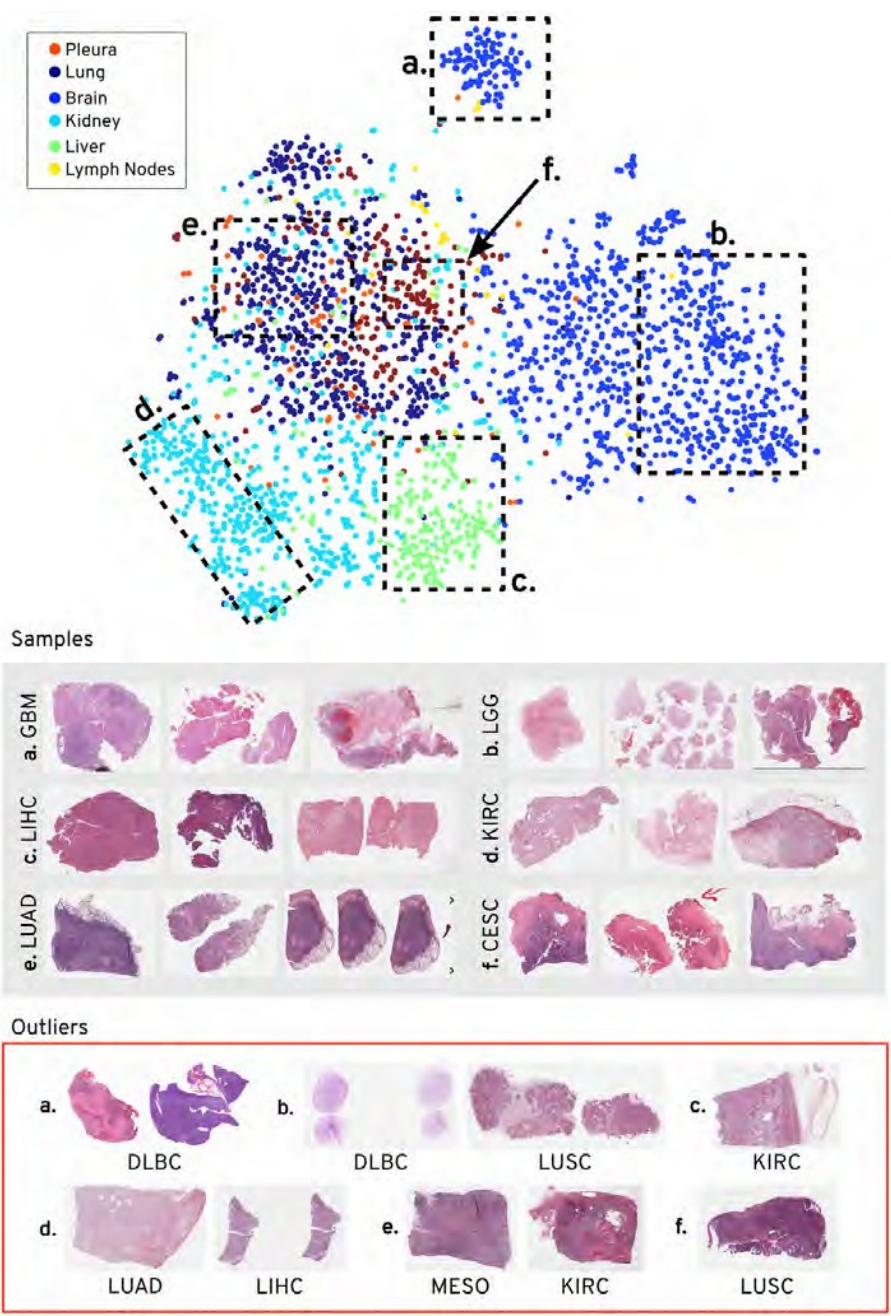

**Figure 9.** T-distributed Stochastic Neighbor Embedding (t-SNE) visualization of pairwise distances of 3000 randomly selected diagnostic slides from six different primary sites. These primary sites are selected to contain top, average, worst accuracy from the Table Table 3—Lung, Brain (top-2), Kidney, Liver (middle-2), Lymph Nodes, and Pleura (bottom-2). Six different areas containing majority of the points from the same cancer subtype are assigned with unique alphabets—**a**, **b**, **c**, **d**, **e**, **f**. The random slides from the majority cancer sub-type within each of the assigned areas are shown in *Samples* box (gray background). The outliers (not belonging to majority the cancer sub-type or the primary site) are shown in the Outliers box (red outline). For example, area a contains majority of scans from Brain with Glioblastoma Multiforme (GBM) whereas its outliers are from Lymph Nodes with Diffuse Large B-cell Lymphoma (DLBC). Without any explicit training, our technique maintains the semantic categories within the diagnostic slides as shows by the t-SNE plot of the pairwise distances. Kidney, Liver, and Brain form different isolated groups whereas lung, pleura, and lymph nodes are intermixed with eachother.